\documentclass[apj]{emulateapj}
\lefthead{TSUJIMOTO \& SHIGEYAMA }
\righthead{THE ORIGINS OF LIGHT AND HEAVY R-PROCESS ELEMENTS}

\def\ltsima{$\; \buildrel < \over \sim\;$}
\def\ltsim{\lower.5ex\hbox{\ltsima}}
\def\gtsima{$\; \buildrel > \over\sim \;$}
\def\gtsim{\lower.5ex\hbox{\gtsima}}
\def\ms{$M_{\odot}$ }
\def\msp{$M_{\odot}$}

\slugcomment{Accepted for publication in ApJ Letters}

\begin{document}
\title{The origins of light and heavy r-process elements identified by chemical tagging of metal-poor stars}

\author{Takuji Tsujimoto\altaffilmark{1} and Toshikazu Shigeyama\altaffilmark{2}}

\affil{$^1$National Astronomical Observatory of Japan, Mitaka-shi,
Tokyo 181-8588, Japan; taku.tsujimoto@nao.ac.jp \\
$^2$Research Center for the Early Universe, Graduate School of Science, University of Tokyo, 7-3-1 Hongo, Bunkyo-ku, Tokyo 113-0033, Japan
}

\begin{abstract}
Growing interests in neutron star (NS) mergers as the origin of r-process elements have sprouted since the discovery of evidence for the ejection of these elements from a short-duration $\gamma$-ray burst.  
The hypothesis of a NS merger origin is reinforced by a theoretical update of nucleosynthesis in NS mergers successful in yielding r-process nuclides with  $A>$130. On the other hand, whether the origin of light r-process elements are associated with nucleosynthesis in NS merger events remains unclear. We find a signature of nucleosynthesis in NS mergers from peculiar chemical abundances of stars belonging to the Galactic globular cluster M15. This finding combined with the recent nucleosynthesis results implies a potential diversity of nucleosynthesis in NS mergers. Based on these considerations, we are successful in the interpretation of an observed correlation between [light r-process/Eu] and [Eu/Fe] among Galactic halo stars and accordingly narrow down the role of supernova nucleosynthesis in the r-process production site. We conclude that the tight correlation by a large fraction of halo stars is attributable to the fact that core-collapse supernovae produce light r-process elements while heavy r-process elements such as Eu and Ba are produced by NS mergers. On the other hand, stars in the outlier, composed of r-enhanced stars ([Eu/Fe]\gtsim+1) such as CS22892-052, were exclusively enriched by matter ejected by a subclass of NS mergers  that is inclined to be massive and consist of both light and heavy r-process nuclides.
\end{abstract}

\keywords{stars: abundances --- ISM: abundances --- Galaxy: evolution --- Galaxy: halo --- stars: neutron}

\section{Introduction}

The recent detection of a near infrared light in the afterglow of a short-duration $\gamma$-ray burst GRB 130603B supports a theoretical prediction that a kilonova as a result of a NS merger \citep{Metzger_10, Metzger_12} is brightened by the radioactivity of very opaque matter composed of almost pure r-process elements \citep{Barnes_13, Tanvir_13, Berger_13,  Hotokezaka_13}.  Realistic calculations for NS mergers performed by several groups thus far have succeeded in producing heavy ($A\sim 130-240$) r-process nuclides \citep[e.g.,][]{Korobkin_12, Bauswein_13} and justified this interpretation. In addition,  \citet{Tsujimoto_14} reexamined the propagation of NS merger ejecta in the interstellar medium (ISM) and found a significant difference from that of supernova ejecta. This finding can resolve a serious problem in the chemical enrichment of r-process elements by NS mergers that has been claimed to be incompatible with observed features of Ba and Eu in the Galactic halo \citep{Mathews_90, Argast_04}. They also constructed a new chemical evolution model of r-process elements, incorporating this finding with the framework in which the Galactic halo is formed through the assembly of numerous protogalactic fragments.

These recent updates are only related to heavy r-process nuclides. Observed chemical abundance patterns of metal-poor halo stars imply that the major source of light r-process elements ($A\approx 90-120$) including Sr, Y, and Zr\footnote{Though the s-process dominates most of these elements in the solar abundance pattern,  the r-process is the major source for chemical enrichment in the Galactic halo. Therefore, light neutron-capture elements are hereafter referred to as light r-process elements.}  is different from that of Eu \citep[][see also Aoki et al. 2005; Fran\c{c}ois et al. 2007]{Montes_07}, and also suggests that the light r-process occurs in the same site of Fe production, i.e., in core-collapse supernovae (CCSNe) \citep{Qian_07}. On the other hand, it is well known that r-enhanced ([Eu/Fe]\gtsim+1) stars \citep[e.g.,][]{Sneden_08} including CS22982-052 \citep{Sneden_03} exhibit the universality of the r-process indistinguishable from the solar r-process abundance pattern, implying a unique site of r-process. At the same time, a deep spectroscopic observation for a bright star HD 122563 reveals a significant enhancement of light r-process element abundances as compared with the solar r-process abundance pattern  \citep{Honda_06}.  From the theoretical side, nucleosynthesis in NS merger is still uncertain in the sense that NS merger produces either only $A>130$ nuclides or all r-process nuclides. Recently proposed models  \citep{Wanajo_14, Just_14} have shown that some NS mergers could produce all r-process nuclides compatible with the observed universal abundance pattern. Thus at the moment, a consensus of identifying the overall sites of r-process including light element primary process (LEPP) has not been achieved yet.

Owing to the high yields of r-process elements per event and the rarity, an NS merger may have a chance to leave its fossil record on stellar abundances. Our study starts with the claim of one compelling candidate for such rare occasions. This is the globular cluster (GC) M15 which hosts stars exhibiting various levels of Eu (and also Ba) enrichment \citep{Sneden_97, Worley_13}. Its observed signature of r-process nucleosynthesis is assessed in the theoretical scheme given by the current nucleosynthesis models of both NS mergers and CCSNe. Then we try to unravel how chemical enrichment of r-process proceeds and makes a correlation between light, heavy r-process elements and Fe among Galactic halo stars, which naturally leads to an identification of the production sites of both light and heavy r-process nuclides. 

\section{Nucleosynthesis signature of a neutron star merger in M15}

Chemical abundance of Galactic metal-poor GC M15 is unusual in the sense that it exhibits a large star-to-star variation in Eu, Ba and La abundances with little scatter in Fe and Ca abundances \citep{Sneden_97, Worley_13}. Abundance patterns of individual stars are confirmed to be of pure r-process origin \citep{Sneden_00}. These facts suggest that heavy r-process enrichment is not associated with CCSNe that provide lighter elements such as Fe and Ca. Alternatively, the hypothesis that a single NS merger occurred after the formation of M15, and polluted the surface of stars by its ejecta well explains a large scatter in the abundance of heavy r-process elements as observed (Shigeyama \& Tsujimoto in preparation). Evolved low-mass stars in the GC had supplied gas with a mass of $\sim1000$ \ms until a NS merger occurred. If the gas is distributed in the core of the GC with a radius of $\sim0.1$ pc, this gas can stop most of r-process elements ejected from the NS merger at speeds of 10-30\% of the speed of light. The resultant [Eu/H] abundance in the gas becomes as high as +0.1 on average. If  a 0.8 \ms star with [Eu/H]=$-2.1$ and velocity dispersion of 15 km s$^{-1}$ accretes this r-process-rich gas at the Bondi accretion rate \citep{Bondi_52} for 10 Myr at minimum, then these stars form a metallicity distribution function (MDF) with respect to [Eu/H] having a peak at [Eu/H]$\sim-1.6$ when they evolve to the red giant branch. Here the accreted matter is assumed to mix with the convective envelope with a mass of 0.3 \msp. The derivation of the formula for the MDF can be found in \citet{Shigeyama_03}. In addition to this peak, we anticipate the second peak at a low [Eu/H] (e.g., [Eu/H]$\sim-2.1$) due to the presence of stars that have not suffered from a pollution by the ejecta of the NS merger. They originally resided outside the core at the NS merger event, and are expected to be distributed over the whole region of the present-day GC owing to two-body relaxation \citep[e.g.,][]{D'Ercole_08}. Such a doubled-peak feature is compatible with the observed results \citep{Worley_13}. 

To host a NS merger event at a rate of about one per 1000-2000 CCSNe \citep{Tsujimoto_14}, a GC should be as massive as several $10^5$ \ms like M15. In addition, a low probability that an NS merger occurs while the GC has a high gas density core should make the presence of Eu-varied GCs quite unique. Indeed, besides M15, only M92 may be another candidate out of all Galactic GCs \citep{Roederer_11} but the Eu variation in M92 is still a question under debate \citep{Cohen_11}.

In contrast to a large variation in heavy r-process abundance among stars, abundances of light neutron-capture elements, Sr, Y, and Zr, do not display any scatter \citep{Otsuki_06, Sobeck_11}. Therefore we conclude that a NS merger in M15 exclusively produced heavy r-process elements.

\section{Production sites of light and heavy r-process elements}

\begin{figure}[t]
\vspace{0.2cm}
\begin{center}
\includegraphics[width=7cm,clip=true]{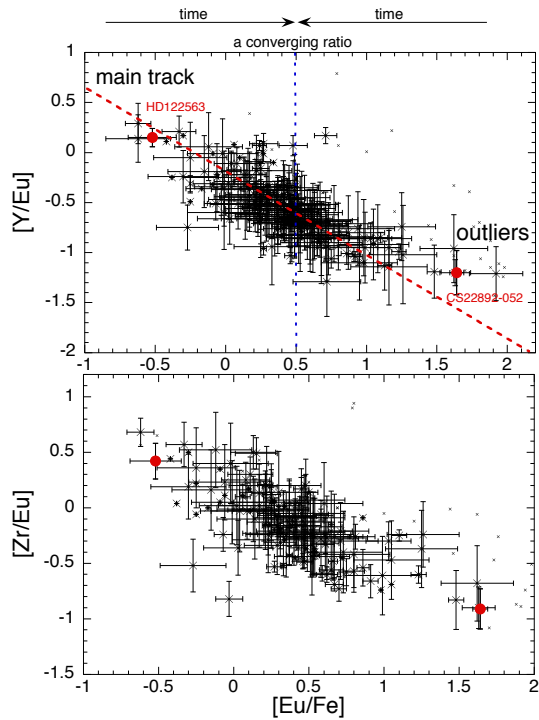}
\end{center}
\vspace{0.3cm}
\caption{Observed correlations of [Y/Eu] (upper panel) and [Zr/Eu] with [Eu/Fe] for Galactic halo stars. The observed Galactic data in the solar vicinity denoted by crosses is selected with [Fe/H]$<-1$ from a database   \citep{Suda_08}. Crosses with error bars are the stars with [Ba/Eu]$<$0 while [Ba/Eu]$\geq$0 by small crosses. We focus on the feature of the stars with [Ba/Eu]$<$0 to avoid a large enrichment by s-process, in particular,  as a result of the transfer of production of s-process elements synthesized in an AGB companion star in a binary to this star. Among them, the observed data for two stars to be highlighted, CS22892-052 and HD122563, are indicated by red circles \citep{Sneden_03, Honda_06}. Brief explanations are attached in the upper panel (see the text). Red dashed line indicates a main track made by a large fraction of stars.
}
\end{figure}

\begin{figure*}[t]
\vspace{0.2cm}
\begin{center}
\includegraphics[width=14cm,clip=true]{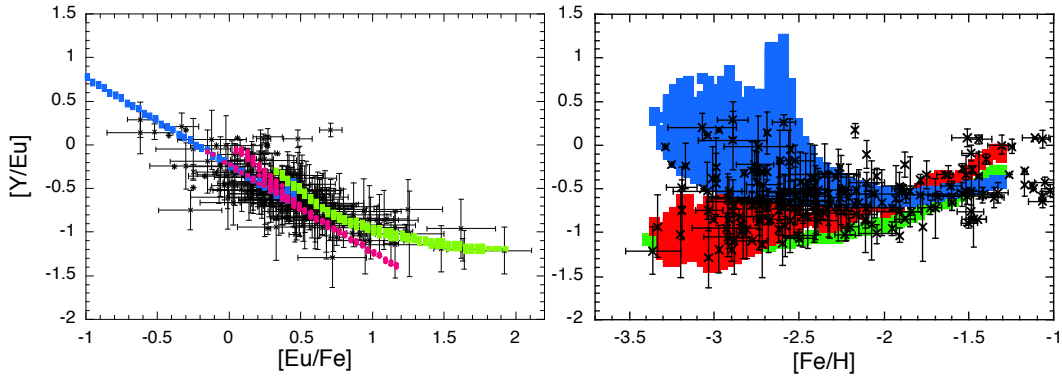}
\end{center}
\vspace{0.3cm}
\caption{Predicted correlations of [Y/Eu] with [Eu/Fe] (left panel) and [Fe/H] (right panel) in the Galaxy halo, compared with the observed data with [Fe/H]$<-1$ and [Ba/Eu]$<$0 \citep{Suda_08}. Our model results are represented by three component with different colors. Most of the observed features are reproduced by blue and red colored tracks. These correspond to stars born from protogalactic fragments with different masses - massive (blue) and small (red) ones. In theses models, all NS mergers are assumed to release only heavy r-process nuclides. On the other hand, green track represents the small-mass fragment hosting one NS merger associated with the production of all r-process nuclides. In all models, CCSNe are assumed to produce light r-process elements.
}
\end{figure*}

Galactic halo stars have a tight correlation of [Sr, Y, Zr/Eu] with [Eu/Fe]  \citep{Montes_07} while the same stars exhibit a large scatter in the [Sr, Y, Zr/Eu] vs. [Fe/H] diagram, which is more frequently used as a diagram displaying chemical features between light and heavy r-process elements. Here we try to unravel the origin of the correlation between [light r-process/Eu] and [Eu/Fe] utilizing the elements Y and Zr since the abundance determination of Sr from the measurement of only a single strong line is less reliable than the other two elements \citep[e.g.,][]{Otsuki_06}. There are three key points to be highlighted in the following theoretical flow; (i) the presence of outlier stars deviating from the tight correlation, (ii) two kinds of nucleosynthesis in NS mergers, and (iii)  different ways of the propagation of ejecta between CCSN and NS merger.

Figure 1 demonstrates that a large fraction of halo stars obey a decreasing trend of [Y(Zr)/Eu] with increasing [Eu/Fe] in the range of $-0.6$\ltsim[Eu/Fe]\ltsim+1.2. This trend is not an end result of a single evolutionary track because in the [Eu/Fe] vs.~[Fe/H] diagram, a large scatter in [Eu/Fe] at the early phase of  [Fe/H]\ltsim$-1.5$  converges to the plateau-like ratio of [Eu/Fe]$\sim+0.5$ in the metallicity range of  $-1.5$\ltsim[Fe/H]\ltsim$-1$ \citep[e.g.,][]{Sneden_08}. Thus, at least two tracks starting from either [Eu/Fe]$\sim -0.6$ or +1.2 and heading for [Eu/Fe]$\sim+0.5$ are required to explain the overall correlation. In addition, some stars deviate from this correlation, in particular all stars with [Eu/Fe]\gtsim +1.2 including CS22892-052,  stay at a ratio of [Y/Eu]$\sim-1.2$ and constitute the outlier. 

First, we discuss the origin of a large contrast of [Y/Eu] in the correlation ranging from $\sim -1.2$ to +0.2. Its contrast can be understood if Y nuclides are supplied from CCSNe while Eu from NS  mergers. This is based on the prediction that the ejecta of a NS merger pervade the whole protogalactic fragment due to their high velocities while CCSNe locally distribute newly synthesized heavy elements only inside the regions swept up by the blast waves \citep{Tsujimoto_14}. This scheme suggests that in a massive (small-mass) fragment, Eu nuclides ejected from a NS merger are mixed with a large (small) amount of ISM  and thus yield a low (high) abundance of Eu while Y nuclides from CCSNe are always mixed with a similar amount  of ISM $\sim5\times10^4$\ms \citep{Shigeyama_98}, resulting in an abundance of Y (also Fe, Mg, etc.) independent of the mass-scales of fragments. Combining these two effects, we can expect that the stellar abundance in a large fragment follows the track initiating from a high [Y/Eu] with a low [Eu/Fe] while the stellar abundance in a small fragment follows the other track associated with a low [Y/Eu] and a high [Eu/Fe]. 

In a small-mass fragment such as $10^6-10^7$ \msp, only one or a few NS merger events are expected to occur. Suppose that one of them is a NS merger synthesizing all r-process nuclides as implied by recent nucleosynthesis results \citep{Wanajo_14, Just_14} and yielding a large amount of Eu. Then stars formed in such a small-mass fragment will have higher [Eu/Fe] ratios and furthermore a non-zero Y yield in the NS merger will upturn [Y/Eu] ratios, resulting in a deviation from the correlation.

In summary, the observed tight correlation between [Y/Eu] and [Eu/Fe] is made by an ensemble of halo stars originating from protogalactic fragments with various mass-scales in which Y and Fe are supplied from CCSNe while Eu from NS mergers.  On the other hand, outlier stars are exclusively enriched by a NS merger, the ejecta of which are massive and include all r-process nuclides. 

\section{Chemical Evolution of [Y/Eu] against [Eu/Fe] and [Fe/H]}

To validate our interpretation presented in the previous section, we model the chemical evolution of the halo and calculate the Y/Eu evolution. The basic model ingredients are the same as in \citet{Tsujimoto_14}. The model represents the chemical evolution for each of gaseous fragments which give birth to halo field stars in the scheme that the halo was formed through accretion of protogalactic fragments.  Fragments with two different masses  are considered, that is, a massive fragment with a mass of $10^9$\ms where NS merger events steadily occur, and the other with $10^7$\ms where only five events are expected.

Our assumptions on NS mergers and CCSNe are as follows. NS mergers occur at a rate of one per 1000 CCSNe with a time delay of 10-30 Myr \citep{Belczynski_06}, and the ejecta are mixed with the entire ISM inside a fragment. A large fraction of NS mergers exclusively release heavy ($A>$130) r-process elements with their  ejecta mass of 0.01 \ms while a minor population produces both light and heavy r-process elements. Their ejected yields are determined to reproduce the observed feature (see below).
On the other hand, CCSNe synthesize light r-process elements together with Fe, and their ejecta are mixed with the ISM locally swept-up by SNe of the mass $5\times10^4$\msp. Here we adopt the Y yields that have a mass dependence similar to Fe in CCSNe. We also consider s-process contribution to Y from massive stars in the metallicity range of [Fe/H]$>-2$ so that the Y yields increase in proportion to the metallicity \citep{Tsujimoto_11}.

The calculated features of [Y/Eu] are shown with respect to [Eu/Fe] as well as [Fe/H] in Figure 2. There are two major evolutionary tracks denoted by blue and red colors. They make a tight correlation between [Y/Eu] and [Eu/Fe] (left panel) and are matched with a large part of the observed data in the [Y/Eu] vs.~[Fe/H] diagram (right panel). These are the results of models for massive ($\sim10^9$\msp;  blue) and $10^7$\ms (red) fragments with NS mergers ejecting only heavy r-process elements. In addition to them, to cover outlier stars in the [Y/Eu]-[Eu/Fe] correlation, we introduce an additional model in which one of five NS mergers releases the ejecta composed of both light and heavy r-process nuclides in a $10^7$\ms fragment. Defining $x$ as Eu yield in units of that of a NS merger ejecting only heavy r-process elements and $y$ as the yield ratio of Y/Eu, our adopted nucleosynthesis yield for this NSM ($x$, $y$) is expressed as (5, 2), and the results are shown by green tracks. It should be of note that though such NS mergers are expected to occur in massive fragments as well, their signature will not become apparent in stellar abundances owing to numerous NS merger events and a high Y abundance in the ISM.

\section{Conclusions}

We have assessed the origin of light and heavy r-process elements from detailed analysis of chemical abundances of Galactic halo field stars together with an implication from chemical feature of the GC M15 and theoretical predictions of nucleosynthesis in NS mergers. Finally, we draw the conclusion that the production site of light r-process elements is multiple; CCSNe are the major source of light r-process elements and a subclass (minority) of NS mergers also release light r-process elements as well as heavy r-process elements. On the other hand, heavy r-process elements have a unique production site - NS mergers -. In other words, there are two types of NS mergers; one type exclusively ejects  heavy r-process nuclides and the other produces both light and heavy r-process nuclides. In this scheme, the universality of r-process pattern seen among r-enhanced stars is reproduced by the abundance pattern of one class of NS mergers that copiously synthesize all r-process nuclides. However, such stars can be regarded as outlier stars, thus we expect that most of halo stars should not follow the universality.

\acknowledgements

The authors wish to thank anonymous referees for their valuable comments which have considerably improved the paper. This paper is based upon work supported in part by the JSPS Grants-in-Aid for Scientific Research (23224004) and under the hospitality of Institute for Nuclear Theory, University of Washington (Report Number: INT-PUB-14-043).

\end{document}